\documentclass[journal]{journal}
\ifCLASSINFOpdf
\else
\fi

\usepackage[utf8]{inputenc} 
\usepackage[T1]{fontenc}    
\usepackage{times}
\usepackage{latexsym}
\usepackage{url}
\usepackage{amsmath}
\usepackage{tabularx}
\usepackage{booktabs}
\usepackage{amssymb}
\usepackage{pifont}
\usepackage{bm}
\usepackage{multirow}
\usepackage{tcolorbox}
\usepackage[framemethod=tikz]{mdframed}
\usepackage{tikz,pgfplots}
\usepgfplotslibrary{groupplots}
  {\begin{enumerate}≈ \footnotesize%
    \setlength{\itemsep}{0pt}%
    \setlength{\parskip}{0pt}%
    }%
  {\end{enumerate}}

  {\begin{itemize}[topsep=0pt, partopsep=0pt] %
    \setlength{\itemsep}{0pt}%
    \setlength{\parskip}{0pt}%
    }%
  {\end{itemize}}

\usepackage{caption}

\usepackage{subcaption}
\usepackage{microtype}
\usepackage{graphicx}
\usepackage{booktabs} 
\usepackage{comment}
\usepackage{bm}
\usepackage{hyperref}
\usepackage{url}
\usepackage{parskip}
\usepackage{amsmath, amsfonts, amsthm}
\usepackage{tikz,pgfplots}

\definecolor{g-red}{HTML}{DB4437}
\definecolor{g-blue}{HTML}{4285F4}
\definecolor{g-green}{HTML}{0F9D58}
\definecolor{g-yellow}{HTML}{F4B400}
\definecolor{g-orange}{HTML}{FF9800}
\definecolor{g-grey}{HTML}{9E9E9E}
\definecolor{shannon}{HTML}{304FFE}
\definecolor{uw}{RGB}{138,43,226}
\definecolor{stanford}{RGB}{255,69,0}
\definecolor{const}{RGB}{68, 110, 182}
\definecolor{head}{RGB}{246, 180, 32}
\definecolor{freq}{RGB}{0, 0, 0}

\begin{document}
%
\title{Analyzing  COVID-19 on Online Social Media: Trends, Sentiments and  Emotions}
%
%
%

\author{~~\\ {\bf Xiaoya Li$^{\clubsuit}$,  Mingxin Zhou$^{\clubsuit}$, Jiawei Wu$^{\clubsuit}$, Arianna Yuan$^{ \blacklozenge} $, Fei Wu$^{\spadesuit}$ and Jiwei Li$^{\clubsuit}$} \\
$^{\spadesuit}$ Department of Computer Science and Technology, Zhejiang University\\
$^\blacklozenge$ Computer Science Department, Stanford University\\
$^{\clubsuit}$ Shannon.AI  ~~\\
   {\{xiaoya\_li, mingxin\_zhou, jiawei\_wu, jiwei\_li\}@shannonai.com} \\xfyuan@stanford.edu, wufei@cs.zju.edu.cn
  }
\markboth{Journal of \LaTeX\ Class Files,~Vol.~6, No.~1, January~2007}%
{Shell \MakeLowercase{\textit{et al.}}: Bare Demo of IEEEtran.cls for Journals}
%



\maketitle
\thispagestyle{empty}

\begin{abstract}
At the time of writing, the ongoing 
pandemic of coronavirus
disease 
 (COVID-19) has caused severe 
impacts on 
 society, economy and people's daily lives. 
People constantly 
 express their
opinions on various aspects  of the pandemic on social media, making 
user-generated content an important  source for understanding 
 public emotions and concerns. 
 
 In this paper,
 we  perform a comprehensive analysis on the affective trajectories of the American people and the Chinese people based on Twitter and Weibo posts between January 20th, 2020 and May 11th 2020. Specifically, by identifying people's sentiments, emotions (i.e., anger, disgust, fear, happiness, sadness, surprise) 
  and the emotional triggers (e.g., what  a user is angry/sad about) 
  we are able to depict the dynamics of public affect in the time of COVID-19. 
  By contrasting two very different countries, China and the Unites States, we reveal sharp differences in people's views on COVID-19 in different cultures. Our study provides a computational approach to unveiling public emotions and concerns on the pandemic in real-time, which would potentially help policy-makers better understand people's need and thus make optimal policy.  
\end{abstract}

\section{Introduction}
The emergence of COVID-19 in early 2020 and its 
subsequent outbreak have affected and changed the world dramatically. 
According to the World Health Organization (WHO), by mid-May 2020, the number of confirmed COVID-19 cases has reached 5 millions with death toll over 300,000 world wide
. 
Several mandatory rules have been introduced by the government to prevent the spread of the coronavirus, such as social distancing, bans on social gatherings, store closures and school closures. Despite their positive effects on slowing the spread of the pandemaic, they neverthless caused severe impacts on the society, the economy and people's everyday life. There have been anti-lockdown and anti-social-distancing protests in many places around the world. Given these difficult situations, it is crucial for policy-makers  to understand people's opinions toward the pandemic so that they can (1) balance the concerns of stoping the pandemic on the one hand and keeping people in good spirits on the other hand and (2) anticipate people's reactions to certain events and policy so that the policy-makers can prepare in advance. More generally, a close look at the public affect during the time of COVID-19 could help us understand people's reaction and thoughts in the face of extreme crisis, which sheds light on humanity in moments of darkness.

People constantly post about the pandemic on social media such as Twitter, Weibo  and Facebook. They express their
attitudes and feelings regarding various aspects of the pandemic, such as the medical treatments, public policy, their worry, etc. Therefore, user-generated content on social media  provides  an important source for understanding 
 public emotions and concerns.   

 In this paper, we provide a comprehensive analysis on the affective trajectories of American people and Chinese people based on Twitter and Weibo posts between January 20th, 2020 and May 11th 2020.
We identify fine-grained emotions (including anger, disgust, fear, happiness, sadness, surprise)  expressed
 on social media based on the user-generated content. 
  Additionally,
  we build NLP taggers to 
   extract the triggers of different emotions, e.g., why people are angry or surprised, what they are worried, etc. 
    We also contrast public emotions  between China and the Unites States, 
 revealing sharp differences in public reactions towards COVID-19 related issues in different countries. 

By tracking the change of public sentiment and emotion  over time, 
our work sheds light on the evolution of public attitude towards this global crisis. 
This work contributes to the growing body of research on social media content in the time of COVID-19.
Our study provides a way to extracting public emotion towards the pandemic in real-time, and could potentially lead to better decision-making and the development of wiser interventions to fight this global crisis.  

The rest of this paper is organized as follows:
we briefly go through some related work in Section 2. 
We then present the analyses on topic trends in Weibo and Twitter (section 3), the extracted emotion trajectories (section 4) and triggers of those emotions (section 5). We finally conclude this paper in Section 6.

\section{Related Work}
\subsection{Analyses on Social Media Content about COVID-19  }

At the time of writing, analyses on people's discussions and behaviors on social media  in the context of COVID-19 has attracted increasing attention. 
\cite{alshaabi2020world} 
analyzed tweets concerning  
COVID-19 on
Twitter by
selecting important 1-grams based on rank-turbulence divergence
and compare languages used in early 2020 with the ones used a year ago. 
The authors observed the first peak of public attention to COVID-19 around January 2020 with the first wave of infections in China, and the 
second peak later when the outbreak hit many western countries. 
\cite{chen2020covid} released the first COVID-19 Twitter dataset. 
\cite{kleinberg2020measuring}
provided a ground truth corpus by annotating 5,000
texts (2,500 short + 2,500 long texts) in UK and
showed
 people's worries about their families and economic situations. 
\cite{gao2020mental} viewed emotions and sentiments on social media as indicators of mental
health issues, which result from self-quarantining and social isolation. 
\cite{schild2020go} revealed increasing amount of hateful speech and conspiracy theories
towards  specific ethnic groups such as Chinese
on Twitter and 4chan’s. 
Other researchers started looking at the spread of misinformation on social media \cite{gallotti2020assessing,ferrara2020covid}.  \cite{cinelli2020covid}
provide an in-depth analysis on the diffusion of misinformation concerning COVID-19 
on five different social platforms. 

\subsection{Analyses of Emotions and Sentiments on Social Media}
{\it Discrete Emotion Theory } \cite{plutchik1980general,frijda1988laws,parrott2001emotions} think that all humans have an innate set of distinct basic emotions. 
Paul Ekman and his colleagues \cite{ekman1992argument} proposed that 
 the six basic emotions of humans  are anger, disgust, fear, happiness, sadness, and surprise.
 Ekman explains that different emotions have particular characteristics expressed in varying degrees. 
 Researchers have debated over the exact categories of discreate emotions. For instance, \cite{raghavan1975number} 
 proposed eight classes for emotions including love, mirth, sorrow, anger, energy, terror, disgust and astonishment.

 Automatically detecting sentiments and emotions in text is a crucial problem in NLP and there has been a large body of
 work on annotating texts based on sentiments and building machine tools to automatically identify emotions
 and sentiments  \cite{yang2007emotion,pang2008opinion,mohammad2016sentiment,tang2015document}. 
\cite{mohammad2017wassa} created the first annotated dataset for four classes of emotions, anger, fear, joy, and sadness, in which 
each text is annotated with not only a label of emotion category, but also the intensity of the emotion expressed based on the Best–Worst Scaling (BWS) technique \cite{louviere1991best}. 
 A follow-up work by \cite{mohammad2018semeval} created a more comprehensively annotated dataset from tweets in English, Arabic, and Spanish. The dataset covers five different sub-tasks including 
 emotion classification, emotion intensity
regression, emotion intensity ordinal classification, valence  regression and valence ordinal classification. 
 
There has been a number of studies on  
extracting 
aggregated public mood and emotions from social media \cite{mishne2006capturing,liu2007arsa,dodds2010measuring,gilbert2010widespread}. 
Facebook introduced Gross National Happiness (GNH)  to estimate the aggregated mood  of the 
public using 
the LIWC dictionary. Results show a clear weekly cycle of public mood. 
\cite{mitchell2013geography} 
and \cite{li2014nasty} 
specially investigate the influence of geographic places and weather on public mood from Twitter data.
The mood indicators extracted from tweets are very predictive and robust \cite{dodds2010measuring,dodds2011temporal}. Therefore, 
they have been used to predict real-world outcomes  
such as economic trends \cite{gilbert2010widespread,rao2012tweetsmart,rao2012using,zhang2011predicting}, stock market \cite{bollen2011twitter,chung2011predicting}, 
influenza outbreak \cite{li2013early},
 and political events \cite{hopkins2010method,o2010tweets,tumasjan2010predicting,larsson2012studying}.

\section{General Trends for COVID-19 Related Posts}
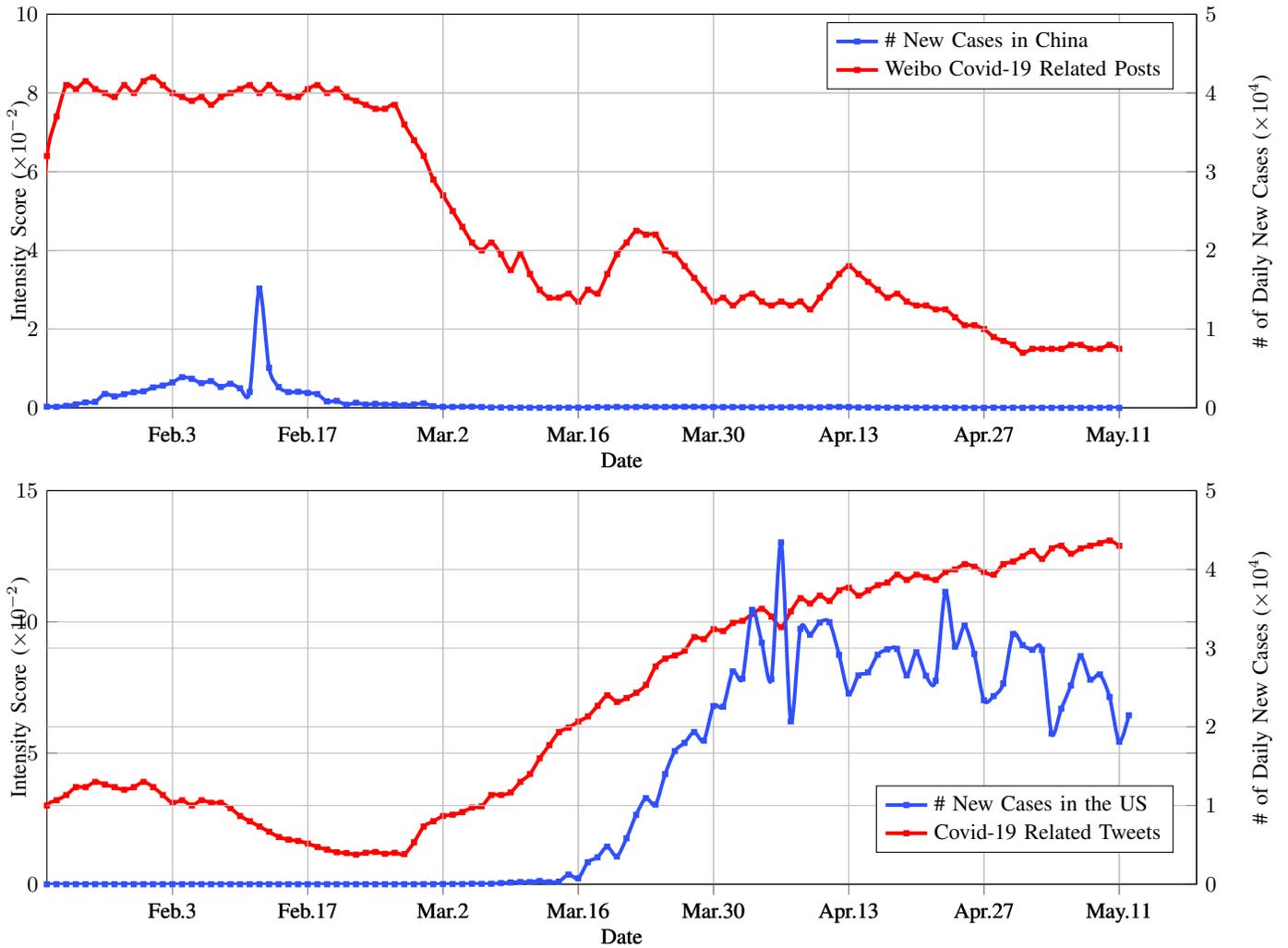
\begin{figure*}[!ht]     
     \centering
     \begin{subfigure}{1\textwidth}
\begin{tikzpicture}
\begin{axis}[
    	   width=1.0\columnwidth,
	    height=0.4\columnwidth,
	    legend cell align=left,
	    xtick={0,14,28,42,56,70,84,98,112},
	    xticklabels={Jan.20, Feb.3, Feb.17, Mar.2, Mar.16, Mar.30, Apr.13, Apr.27, May.11},
	    xticklabel style = {font=\small},
   		ymin=0, ymax=10,
   		xtick pos=left,
   		xtick align=outside,
	    xmin=1,xmax=120,
	    mark options={mark size=0.5},
		font=\small,
   	 	ymajorgrids=true,
    	xmajorgrids=true,
    	xlabel=Date,
        ylabel=Intensity Score  ($\times 10^{-2}$),
    	ylabel style={at={(axis description cs: 0.05, 0.5)}}
	]
    	
\addplot[
    smooth,
    color=red,
    mark=square*,
    line width=1.5pt
    ]
    coordinates {
(0,3.4)(1,6.4)(2,7.4)(3,8.2)(4,8.1)(5,8.3)(6,8.1)(7,8.0)(8,7.9)(9,8.2)(10,8.0)(11,8.3)(12,8.4)(13,8.2)(14,8.0)(15,7.9)(16,7.8)(17,7.9)(18,7.7)(19,7.9)(20,8.0)(21,8.1)(22,8.2)(23,8.0)(24,8.2)(25,8.0)(26,7.9)(27,7.9)(28,8.1)(29,8.2)(30,8.0)(31,8.1)(32,7.9)(33,7.8)(34,7.7)(35,7.6)(36,7.6)(37,7.7)(38,7.2)(39,6.8)(40,6.4)(41,5.8)(42,5.4)(43,5.0)(44,4.6)(45,4.2)(46,4.0)(47,4.2)(48,3.9)(49,3.5)(50,3.9)(51,3.4)(52,3.0)(53,2.8)(54,2.8)(55,2.9)(56,2.7)(57,3.0)(58,2.9)(59,3.4)(60,3.9)(61,4.2)(62,4.5)(63,4.4)(64,4.4)(65,4.0)(66,3.9)(67,3.6)(68,3.3)(69,3.0)(70,2.7)(71,2.8)(72,2.6)(73,2.8)(74,2.9)(75,2.7)(76,2.6)(77,2.7)(78,2.6)(79,2.7)(80,2.5)(81,2.8)(82,3.1)(83,3.4)(84,3.6)(85,3.4)(86,3.2)(87,3.0)(88,2.8)(89,2.9)(90,2.7)(91,2.6)(92,2.6)(93,2.5)(94,2.5)(95,2.3)(96,2.1)(97,2.1)(98,2.0)(99,1.8)(100,1.7)(101,1.6)(102,1.4)(103,1.5)(104,1.5)(105,1.5)(106,1.5)(107,1.6)(108,1.6)(109,1.5)(110,1.5)(111,1.6)(112,1.5)
    };\label{plot_1_y1}
\end{axis}

\begin{axis}[
 	   axis y line*=right,
    	   width=1.0\columnwidth,
	    height=0.4\columnwidth,
	    legend cell align=left,
	    xtick={0,14,28,42,56,70,84,98,112},
	    xticklabels={Jan.20, Feb.3, Feb.17, Mar.2, Mar.16, Mar.30, Apr.13, Apr.27, May.11},
	    xticklabel style = {font=\small},
   		ymin=0, ymax=5,
   		xtick pos=left,
   		xtick align=outside,
	    xmin=1,xmax=120,
	    mark options={mark size=0.5},
		font=\small,
   	 	ymajorgrids=true,
    	xmajorgrids=true,
    	xlabel=Date,
        ylabel=\# of Daily New  Cases  ($\times 10^{4}$),
    	ylabel style={at={(axis description cs: 1.13, 0.5)}}
	]
    	
\addplot[
    smooth,
    color=shannon,
    mark=square*,
    line width=1.5pt
    ]
    coordinates {
(0,0.0077)(1,0.0149)(2,0.0131)(3,0.0259)(4,0.0444)(5,0.0688)(6,0.0769)(7,0.1771)(8,0.1459)(9,0.1737)(10,0.1982)(11,0.2102)(12,0.259)(13,0.2829)(14,0.3235)(15,0.3893)(16,0.3697)(17,0.3143)(18,0.3401)(19,0.2656)(20,0.3062)(21,0.2484)(22,0.2022)(23,1.5153)(24,0.5093)(25,0.2644)(26,0.2009)(27,0.2051)(28,0.1891)(29,0.1751)(30,0.082)(31,0.0892)(32,0.0399)(33,0.0649)(34,0.0416)(35,0.0517)(36,0.0411)(37,0.044)(38,0.0329)(39,0.043)(40,0.0579)(41,0.0206)(42,0.0128)(43,0.012)(44,0.0143)(45,0.0145)(46,0.0103)(47,0.0046)(48,0.0045)(49,0.002)(50,0.0031)(51,0.0025)(52,0.0011)(53,0.0018)(54,0.0027)(55,0.0029)(56,0.0039)(57,0.0035)(58,0.0084)(59,0.0065)(60,0.0116)(61,0.0082)(62,0.0102)(63,0.0147)(64,0.0099)(65,0.0114)(66,0.0118)(67,0.0135)(68,0.0128)(69,0.0106)(70,0.0098)(71,0.0086)(72,0.0093)(73,0.0078)(74,0.0073)(75,0.0055)(76,0.0075)(77,0.0066)(78,0.0086)(79,0.0092)(80,0.0056)(81,0.0064)(82,0.0113)(83,0.0115)(84,0.0099)(85,0.0049)(86,0.0052)(87,0.0027)(88,0.0031)(89,0.0021)(90,0.0036)(91,0.0013)(92,0.0037)(93,0.0015)(94,0.0009)(95,0.0013)(96,0.0014)(97,0.0003)(98,0.0006)(99,0.0022)(100,0.0004)(101,0.0012)(102,0.0003)(103,0.0005)(104,0.0007)(105,0.0004)(106,0.0002)(107,0.0003)(108,0.0006)(109,0.0001)(110,0.0014)(111,0.002)(112,0.0001)

    };\label{plot_1_y2}
        
     \addlegendentry{\#  New Cases in China}

    \addlegendimage{/pgfplots/refstyle=plot_1_y1}    
    \addlegendentry{Weibo Covid-19 Related Posts}

\end{axis}

\end{tikzpicture}
     \end{subfigure}

              \label{general}

                   \begin{subfigure}{1\textwidth}
\begin{tikzpicture}
\begin{axis}[
    	   width=1.0\columnwidth,
	    height=0.4\columnwidth,
	    legend cell align=left,
	    legend style={at={(1, 0.5)},anchor=south east},
	    xtick={0,14,28,42,56,70,84,98,112},
	    xticklabels={Jan.20, Feb.3, Feb.17, Mar.2, Mar.16, Mar.30, Apr.13, Apr.27, May.11},
	    xticklabel style = {font=\small},
   		ymin=0, ymax=15,
   		xtick pos=left,
   		xtick align=outside,
	    xmin=1,xmax=120,
	    mark options={mark size=0.5},
		font=\small,
   	 	ymajorgrids=true,
    	xmajorgrids=true,
    	xlabel=Date,
        ylabel=Intensity Score  ($\times 10^{-2}$),
    	ylabel style={at={(axis description cs: 0.05, 0.5)}}
	]

\addplot[
    smooth,
    color=red,
    mark=square*,
    line width=1.5pt
    ]
    coordinates {
(0,2.0)(1,3.0)(2,3.2)(3,3.4)(4,3.7)(5,3.7)(6,3.9)(7,3.8)(8,3.7)(9,3.6)(10,3.7)(11,3.9)(12,3.7)(13,3.4)(14,3.1)(15,3.2)(16,3.0)(17,3.2)(18,3.1)(19,3.1)(20,2.9)(21,2.6)(22,2.4)(23,2.2)(24,2.0)(25,1.8)(26,1.7)(27,1.65)(28,1.55)(29,1.42)(30,1.32)(31,1.22)(32,1.19)(33,1.13)(34,1.2)(35,1.23)(36,1.16)(37,1.2)(38,1.15)(39,1.6)(40,2.2)(41,2.4)(42,2.6)(43,2.65)(44,2.75)(45,2.92)(46,2.97)(47,3.4)(48,3.4)(49,3.5)(50,3.9)(51,4.2)(52,4.8)(53,5.3)(54,5.8)(55,5.97)(56,6.2)(57,6.4)(58,6.8)(59,7.2)(60,6.95)(61,7.1)(62,7.3)(63,7.6)(64,8.3)(65,8.6)(66,8.72)(67,8.9)(68,9.42)(69,9.34)(70,9.72)(71,9.65)(72,9.96)(73,10.04)(74,10.3)(75,10.5)(76,10.2)(77,9.8)(78,10.4)(79,10.9)(80,10.7)(81,11.0)(82,10.8)(83,11.2)(84,11.3)(85,11.0)(86,11.2)(87,11.4)(88,11.5)(89,11.8)(90,11.6)(91,11.8)(92,11.7)(93,11.6)(94,11.9)(95,12.0)(96,12.2)(97,12.1)(98,11.9)(99,11.8)(100,12.2)(101,12.3)(102,12.5)(103,12.7)(104,12.4)(105,12.8)(106,12.9)(107,12.6)(108,12.8)(109,12.9)(110,13.0)(111,13.1)(112,12.9)
    };\label{plot_1_y1}

\end{axis}

\begin{axis}[
 	   axis y line*=right,
    	   width=1.0\columnwidth,
	    height=0.4\columnwidth,
	    legend cell align=left,
	    legend style={at={(0.98, 0.08)},anchor=south east},
	    xtick={0,14,28,42,56,70,84,98,112},
	    xticklabels={Jan.20, Feb.3, Feb.17, Mar.2, Mar.16, Mar.30, Apr.13, Apr.27, May.11},
	    xticklabel style = {font=\small},
   		ymin=0, ymax=5,
   		xtick pos=left,
   		xtick align=outside,
	    xmin=1,xmax=120,
	    mark options={mark size=0.5},
		font=\small,
   	 	ymajorgrids=true,
    	xmajorgrids=true,
    	xlabel=Date,
        ylabel=\# of Daily New  Cases  ($\times 10^{4}$),
    	ylabel style={at={(axis description cs: 1.13, 0.5)}}
	]

\addplot[
    smooth,
    color=shannon,
    mark=square*,
    line width=1.5pt
    ]
    coordinates {
(0,0.0)(1,0.0)(2,0.0)(3,0.0)(4,0.0001)(5,0.0)(6,0.0003)(7,0.0)(8,0.0)(9,0.0)(10,0.0)(11,0.0002)(12,0.0001)(13,0.0)(14,0.0003)(15,0.0)(16,0.0)(17,0.0)(18,0.0)(19,0.0)(20,0.0)(21,0.0)(22,0.0001)(23,0.0)(24,0.0001)(25,0.0)(26,0.0)(27,0.0)(28,0.0)(29,0.0)(30,0.0)(31,0.0)(32,0.0002)(33,0.0)(34,0.0)(35,0.0)(36,0.0)(37,0.0)(38,0.0001)(39,0.0)(40,0.0008)(41,0.0006)(42,0.0023)(43,0.0025)(44,0.002)(45,0.0066)(46,0.0047)(47,0.0064)(48,0.0147)(49,0.0225)(50,0.029)(51,0.0278)(52,0.0414)(53,0.0267)(54,0.0338)(55,0.1237)(56,0.0755)(57,0.2797)(58,0.3419)(59,0.4777)(60,0.3528)(61,0.5836)(62,0.8821)(63,1.0934)(64,1.0115)(65,1.3987)(66,1.6916)(67,1.7965)(68,1.9332)(69,1.8251)(70,2.2635)(71,2.2562)(72,2.7043)(73,2.6135)(74,3.4864)(75,3.0683)(76,2.6065)(77,4.3438)(78,2.0682)(79,3.2448)(80,3.1705)(81,3.3251)(82,3.3288)(83,2.9145)(84,2.4242)(85,2.6527)(86,2.693)(87,2.9164)(88,2.9836)(89,2.9895)(90,2.6543)(91,2.9468)(92,2.649)(93,2.5858)(94,3.7144)(95,3.0181)(96,3.2853)(97,2.9256)(98,2.3371)(99,2.3901)(100,2.5512)(101,3.1787)(102,3.0369)(103,2.9794)(104,2.9763)(105,1.9138)(106,2.2303)(107,2.5253)(108,2.8974)(109,2.5996)(110,2.666)(111,2.3792)(112,1.8106)(113,2.1467)
    };\label{plot_1_y2}
    
    \addlegendentry{\#  New Cases in the US }

        \addlegendimage{/pgfplots/refstyle=plot_1_y1} 
    \addlegendentry{Covid-19 Related Tweets}

\end{axis}

\end{tikzpicture}

     \end{subfigure}
              \caption{(a) Daily intensity scores for Covid-19 topics on Weibo and the number of daily  cases reported by Chinese CDC. 
               (a) Daily intensity scores for Covid-19 topics on Twitter and the number of daily  cases reported by US CDC. 
               Intensity scores are  the number Covid related posts
divided by the total number of retrieved daily posts. }

              \label{general}
\end{figure*}
In this section, 
we present the general trends for COVID19-related posts  
on Twitter and Weibo. 
We first present the semi-supervised models we used to detect COVID-19 related tweets.
Next we present the analysis on the topic trends on 
 the two social media platforms.  

\subsection{Retrieving COVID-19 Related Posts}
For Twitter, we first obtained 1\% of all tweets that are written in English and published within the time period between January 20th, 2020 and May 11th 2020.  
The next step is to select tweets related to COVID-19.
The simplest way, as in \cite{chen2020covid,ferrara2020covid}, is to use a handcrafted keyword list to obtain tweets containing words found in the list. 
However, 
 this  method  leads to lower values in both precision and recall: 
for precision, user-generated content that contains the  mention of a keyword is not necessarily related to COVID-19. 
For example, the keyword list used in \cite{chen2020covid} include the word {\it China}, and it is not suprising that a big proportion of the posts containing ``China" is not related to COVID-19; for recall, keywords for COVID-19 can change over time and might be missing in the keyword list. 

To tackle this issue, 
we  adopt a  bootstrapping 
approach. The 
bootstrapping approach is related to previous work
on semi-supervised data harvesting methods \cite{li2014weakly,davidov2007fully,kozareva2010learning},  in which 
we build 
 a model that 
 recursively uses seed examples
to extract patterns, which are then used to harvest
new examples. Those new examples are further used as new
seeds to get new patterns. 
To be specific, we first obtained a starting seed keyword list by (1) ranking  words 
based on  tf-idf scores 
from eight COVID-19 related wikipedia articles; (2) 
manually examining the ranked word list, removing those words that are apparently not COVID-19 related, and use the top 100 words in the remaining items. 
Then we retrieved tweets with the mention of those keywords. 
Next, we randomly sampled 1,000 tweets from the collection and manually labeled
them as either COVID-19 related or not.
The labeled dataset is split into the training, development and test sets with ratio 8:1:1. 
 A binary classification model is trained on the labeled dataset 
 to classify whether a post with the mention of COVID-related keywords is actually COVID-related. 
 The model 
 is trained using BERT \cite{devlin2018bert} and optimized using Adam \cite{kingma2014adam}. 
  Hyperparameters such as the batch size, learning rate are tuned on the development set. 
  Next, we obtain a new seed list by picking the most salient words that contribute to the positive category in the binary classification model
  based on the first-order derivative saliency scores  \cite{erhan2009visualizing,simonyan2013deep,li2015visualizing}. 
  This marks the end of the first round of the bootstrapping. 
  Next we used the new keyword list to re-harvest a new dataset with the mention of the keyword, 1,000 of which is selected and labeled to retrain the binary classification model. 
 We repeat this process for three times.
 F1 scores for the three rounds of binary classification are 0.74, 0.82, 0.86 respectively. 
After the final round of bootstrapping,
 we collected a total number of 78 million English tweets concerning the topic of COVID-19.  
We used this strategy to retrieve COVID-related posts on Weibo and collected a total number of 16 million posts.  

\subsection{Analyses}
We report the intensity scores for Weibo and Twitter in Figure  \ref{general}.
We split all tweets by date, where $X_t$ denotes all tweets published on day $t$.  
The value of intensity is the number of posts classified as  COVID-related 
divided by the total number of retrieved posts, i.e., $|X_t|$.  
On Weibo, we observe a peak  in late January and  February, then a drop, followed by another  rise in March, and a gradual decline afterwards. 
The trend on Chinese social media largely reflects the progress of the pandemic in China: 
the outbreak of COVID-19 and the  spread from Wuhan to the rest of the country corresponds to the first peak.
 The subsequent drop reflects the promise in containing the virus, followed by a minor relapse in March. 
For Twitter, we observe a small peak that is aligned
with the news from China about the virus. 
The subsequent drop reflects the decline of the attention to the outbreak in China. The curve progressively went up since March, corresponding to the outbreak in the US. 
Upon the writing of this paper, we have not observed a sign of drop in the intensity score of COVID19-related posts.

\section{The Evolution of Public Emotion}
In this section, we present the analyses on the evolution of public emotion in the time of COVID-19.
We first present the algorithms we used to identify the emotions expressed in a given post. Next we present the results of the analyses. 

\subsection{Emotion Classification}
We adopted the well-established emotion theory by Paul Ekman \cite{ekman1992argument}, which groups human emotions into 6 major categories, 
i.e., anger, disgust, worry, happiness, sadness, and surprise.
Given a post from a
social network
 user, we assign one or multiple emotion labels to it \cite{buechel2016emotion,demszky2020goemotions}.
This setup is quite common in text classification \cite{ikonomakis2005text,aggarwal2012survey,zhang2015character,zhou2015c,joulin2016bag}. 

For emotion classification of English tweets, we 
take the advantage of labeled datasets from 
 the SemEval-2018 Task 1e \cite{mohammad2018semeval}, in which a tweet  was associated with either 
the ``neutral'' label or with one or multiple emotion
labels by human evaluators. 
The SemEval-2018 Task 1e contains eleven emotion categories in total, i.e., anger, anticipation, disgust, fear, joy, love, optimism, pessimism, 
sadness, surprise and trust, and we only use the datasets for a six-way classification, i.e., 
anger, disgust, fear, happiness, sadness, and surprise. 
Given that 
the domain of the dataset used in \cite{mohammad2018semeval} covers all kinds of tweets, and our domain of research covers only COVID-related tweets, there is a gap between the two domains.
Therefore, we additionally labeled 15k COVID-related tweets following the guidelines in \cite{mohammad2018semeval}, where each tweet can take either the neural label or one/multiple emotion
labels. 
Since one tweet can take on multiple emotion labels, 
the task is formalized as a 
 a multi-label classification task, in which six binary (one vs. the rest) classifiers are trained. 
 We used the description-based BERT model \cite{chai2020description} as the backbone, which achieves current SOTA performances on a wide variety of text classification tasks.
More formally, let us consider 
a to-be-classified tweet 
 $x=\{x_1,\cdots,x_L\}$, where $L$ denotes the length of the text $\bm{x}$. 
 Each $x$ will be tagged with one or more class labels $y \in \mathcal{Y} = [1,N]$, where $N=6$ denotes the number of the predefined emotion classes (the six emotion categories). To compute the probability $p(y|\bm{x})$, each input text $\bm{x}$ is concatenated with the description $\bm{q}_y$ to generate $\{\text{[CLS]};\bm{q}_y;\text{[SEP]};\bm{x}\}$, where $\text{[CLS]}$ and $\text{[SEP]}$ are special tokens.
The description $q_y$ is  the Wikipedia description for each of the emotions. For example, $q_y$ for the category {\it anger} is {\it ``Anger, also known as wrath or rage, is an intense emotional state involving a strong uncomfortable and hostile response to a perceived provocation, hurt or threat."}
 Next, the concatenated sequence
 is fed to  the BERT model, from which we obtain the contextual representations $h_\text{[CLS]}$.
 $h_\text{[CLS]}$ is then transformed to a real value between 0 and 1 using the sigmoid function, representing the probability of assigning
 the emotion
 label $y$ to the input tweet $\bm{x}$:
\begin{equation}
p(y|\bm{x})=\text{sigmoid}(W_2\text{ReLU}(W_1h_\text{[CLS]}+b_1)+b_2)
\end{equation}
where $W_1,W_2,b_1,b_2$ are some parameters to optimize. 
Classification performances for different models are presented in 
 Table \ref{en-emotion}. 
 
 \begin{table}
 \center
 \begin{tabular}{cccc}\hline
 Model & Acc & micro F1 &   macro F1 \\\hline
 SVM biagram & 51.4 & 63.0 & 52.7  \\
 BERT \cite{devlin2018bert} & 65.0 & 75.2 & 66.1 \\
BERT-description  \cite{chai2020description} & 66.8 & 77.0 & 68.3  \\\hline
 \end{tabular} 
\caption{ Results for the multi-label emotion classification for English tweets. }
\label{en-emotion}
 \end{table}

\subsection{Analyses}
For emotion $y$, its intensity score $S(t,y)$ for day $t$ is the average probability (denoted by $P(y|x)$)
of assigning label $y$ to all
the texts in that day $X_t$. For non COVID-related texts,  $P(y|x)$ is automatically set to 0.  We thus have:
\begin{equation}
S(t,y) = \frac{1}{|X_t|} \sum_{x\in X_t} p (y|x)
\end{equation}
For Chinese emotion classification, we used the labeled dataset in \cite{lai2019fine}, which contains 15k labeled microblogs from weibo\footnote{The original dataset contains 7 categories of emotions, and we used only six of them.}. 
In addition to the dataset provided by \cite{lai2019fine}, we labeled COVID-related 20k microblogs. 
The combined dataset is then used to train a multi-label classification model based on the description-BERT model \cite{chai2020description}.
Everyday emotion scores for Weibo are computed in the same way as for Twitter.

The time series of intensity scores of six different emotions, i.e., sadness, anger, disgust, worry, happiness, surprise, 
for Weibo and Twitter 
 are  shown in Figures \ref{ch-emotion} and \ref{en-emotion}, respectively. 
For Weibo,
as can be seen, the trend of  {\it worry} is largely in line with the trend of the general intensity of the COVID-related posts. It reached a peak in late January, and then gradually went down, followed by a small relapse in mid-March. 
For {\it anger}, the intensity first went up steeply at the initial stage of the outbreak, staying high for two weeks, 
and then had another sharp increase 
 around February 8th. The peak on February 8th was due to the death of Wenliang Li, 
a Chinese ophthalmologist 
who issued the  warnings about the virus. 
The intensity for anger then gradually decreased,  with no relapse afterwards. 
The intensity for 
{\it disgust} remained relatively low across time. 
For {\it sadness}, the intensity reached the peak at the early stage of the outbreak, then gradually died out with no 
relapse. 
For {\it surprise}, it went up first, mostly because of the fact that the public was surprised by the new virus and the unexpected outbreak, but then gradually went down.
The intensity for {\it happiness} remained relatively low across time, with a small peak in late April, mostly because the countrywide lockdown was over.

For Twitter, the intensity for {\it worry} went up shortly in late January, followed by a drop. 
The intensity then went up steeply in mid-March in response to the pandemic breakout in the States, reaching a peak around March 20th, then decreased a little bit and remained steady afterwards. 
The  intensity for {\it anger} kept going up after the outbreak in mid-March, with no drop observed. 
The trend for {\it sadness} is mostly similar to that of the overall intensity. 
 For {\it surprise}, the curve went up first after the breakout in early March, reaching a peak around Mar 20th, then dropped, and remained steady afterwards.
 For {\it happiness}, the intensity remained low over time. 
 
\begin{table*}
\small
\center
\begin{tabular}{|c|c|c|c|c|c|}\hline
sadness & anger &  disgust &worry &happiness & surprise \\\hline
passed away &realdonaldtrump   &  covid / covid-19 &covid / covid-19   & healthy &covid / covid-19 \\
\multirow{2}{*}{died}  &  \multirow{2}{*}{lockdown}   & \multirow{2}{*}{realdonaldtrump}& \multirow{2}{*}{job} &  \multirow{2}{*}{help}  &human-to-human \\
& & & &&transmission \\
deaths  & government & chinese &kids  &recover  &outbreak \\
fever  &quarantine& trump&bill& check &  lockdown\\
unemployed & wuhan& masks & unemployed& return to work& test\\ 
parents& lies & virus  &  food &reopening& deaths \\
test positive &  close & pence &    crisis & vaccine &pandemic\\
family &  stayhome & chinks &    parents & money &confirmed cases\\
patients &  WHO & china &    economy & treatment &total numbers\\
isolation & trump & hospitals & families & friends & conspiracy \\\hline
\end{tabular} 
\caption{Top 10 extracted trigger spans regarding different emotions on Twitter.}
\label{top-entity}
\end{table*}

\begin{table*}
\small
\center
\begin{tabular}{|c|c|c|c|c|}\hline
China& lockdown &Trump   & Hospitals &Increasing Cases and Deaths \\\hline
China& quarantine& realdonaldtrump  & hospital& case \\
Chinese& stay & donald & patients & deaths\\
Wuhan& close&lies  & test & report\\
bat-eating&home &republicans  &doctor & confirmed\\
chink& stayhome & hydroxychloroquin  & case & report\\
chinesevirus & boarder & american & healthcare& york \\
sinophobia&shutdown&media & drug &us \\
chingchong & distancing &governor  & vaccine & total \\
hubei& coronalockdown&pence & ventilator & government\\ \hline
\end{tabular}
\caption{Top  mentions of different subcategories for {\it anger} on Twitter.}
\label{anger}
\end{table*}

\begin{table*}
\small
\center
\begin{tabular}{|c|c|c|c|c|}\hline
syndrome and being infected& families &finance and economy  & jobs and food &Increasing  Cases and Deaths \\\hline
fever& parents&  money & jobs &  deaths \\
hospital & mother  & stock & unemployed& spread\\
cough& children&financial &food &poll number \\
test positive&mom &price & money & death toll \\
icu& families & loan &work  & confirmed \\
doctor &father  &business& starve & rise\\
bed&kids& debt & unemployment  & official \\
confimed & daughter &market & check & number \\
sick& father-in-law&  crash& layoff & safe\\ \hline
\end{tabular}
\caption{Top  mentions of different subcategories for {\it worry} anger on Twitter.}
\label{worry}
\end{table*}
\input{figs/ch_emotion.tex}
\input{figs/en_emotion.tex}

 \section{ Emotional Triggers}
For a given emotion, we would like to dive deeper into its different
subcategories. For example, for {\it worry}, we wish to know 
what the public is worried about, and how these triggers of worry change over time.
In this section, we first present our methods for extracting triggers/subcategories for different emotions, followed by 
some analyses and visualization on the Twitter data. 

\subsection{Extracting the Triggers for Different Emotions}
In order to extract the emotional triggers from Twitter’s noisy
text, we first annotate a corpus of tweets. 
For the ease of annotation, each emotion is associated with only a single trigger:
the person/entity/event that a
user has a specific emotion towards/with/about. 
A few examples are shown as follows with target triggers surrounded by brackets.
\begin{itemize}
\item 
{\it Angry protesters are traveling 100's of miles to join organized rallies over COVID-19 [lockdown]$_{attr\_anger}$. }  
\item 
{\it  Feeling awfully tired after a 5.30am start for work today. Worried too about [the early return to school]$_{attr\_worry}$, my grandchildren are so very dear to me . I could not bear to lose them to covid. }
\item 
{\it All Americans are very angry with 
[@realDonaldTrump]$_{attr\_anger}$
 81,647 dead Americans would be very angry as well. If they weren’t dead.}
\item 
{\it 
Fucking  [bat-eating chinks]$_{{attr\_anger}}$, go die in a hole, far away from us. }
\item
{\it Well, I am ANGRY as hell at Trump$_{attr\_anger }$}.
\item 
{\it With great sadness we report the sad [loss of two dear Friends]$_{{attr\_sadness}}$ }
\item 
{\it 
The lockdown$_{{attr\_anger}}$ was implemented when there were hardly any cases and now it is above lakhs and people are acting so carelessly. Making me so angry. }
\end{itemize}

In order to build an emotional trigger tagger, we annotated 2,000 tweets in total, and split them into training, development and test sets with ratio 8:1:1. 
We treat the problem as a sequence labeling task, using Conditional Random Fields for learning and inference with 
 BERT-MRC  features \cite{li2019unified}. 
Comparing with vanilla BERT tagger \cite{devlin2018bert}, 
the BERT-MRC tagger has the strength of encoding the description of the to-be-extracted entities, e.g., {\it what they are worried about}.
As this description provides the prior knowledge about the entities, it has been shown to outperform vanilla BERT even when less training data is used.  
In addition to the representation features from BERT-MRC, we
 also considered the Twitter-tuned POS features \cite{ritter2011named},
the dependency features 
from a Twitter-tuned dependency parsing model  \cite{kong2014dependency} and the Twitter event features \cite{ritter2012open}. 
The precision and recall for segmenting emotional triggers on English tweets are reported in Table \ref{trigger}.

\begin{table*}[!ht]
\center
\begin{tabular}{lccc}\hline
Model & Pre & Rec & F1 \\\hline
BERT  &  0.41 & 0.60  & 0.48\\
BERT-MRC & 0.54 & 0.66 &0.59 \\
CRF with BERT-MRC features & 0.53 &0.68&0.60\\
CRF with BERT-MRC features, POS, event and parse tree features & 0.58 & 0.78 & 0.66 \\\hline
\end{tabular} 
\caption{Performances of different models on emotional trigger extraction from Tweets.}
\label{trigger}
\end{table*}
The precision and recall for segmenting triggering event phrases are
reported in Table 3. We observe a significant performance boost with linguistic features such as POS and dependency features. This is mainly due to the small size of the labeled dataset. The best model achieves an F1 score of  0.66.

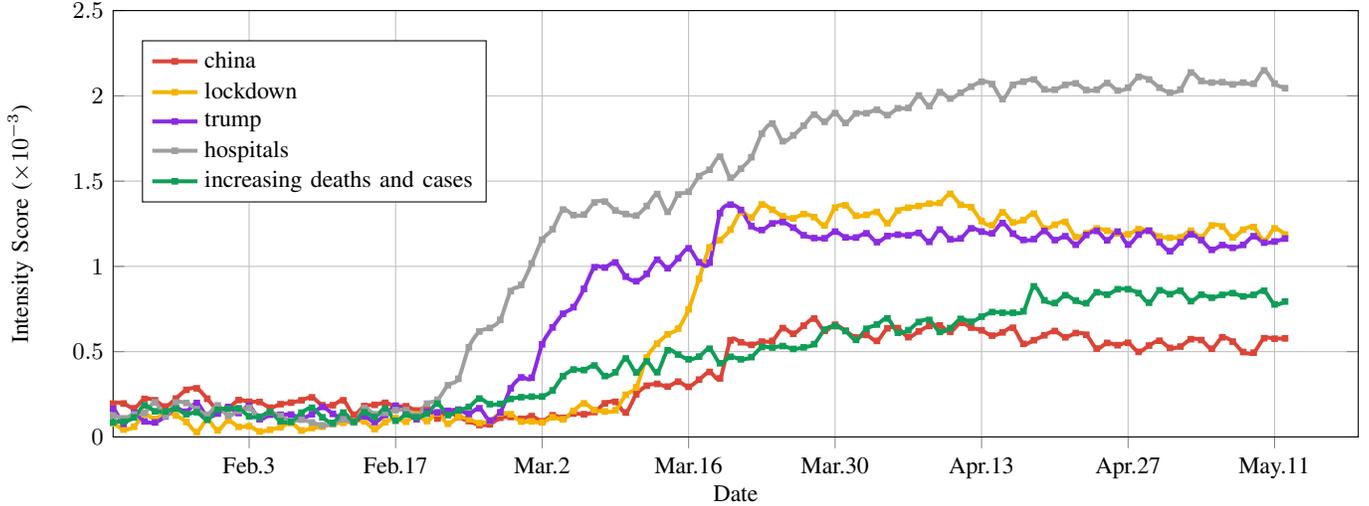
\begin{figure*}[!ht]     
     \centering
     \begin{subfigure}{1\textwidth}
\begin{tikzpicture}
\begin{axis}[
    	   width=1\columnwidth,
	    height=0.4\columnwidth,
	    legend cell align=left,
	    legend style={at={(0.3, 0.55)},anchor=south east},
	    xtick={0,14,28,42,56,70,84,98,112},
	    xticklabels={Jan.20, Feb.3, Feb.17, Mar.2, Mar.16, Mar.30, Apr.13, Apr.27, May.11},
	    xticklabel style = {font=\small},
   		ymin=0, ymax=2.5,
   		xtick pos=left,
   		xtick align=outside,
	    xmin=1,xmax=120,
	    mark options={mark size=0.5},
		font=\small,
   	 	ymajorgrids=true,
    	xmajorgrids=true,
    	xlabel=Date,
        ylabel=Intensity Score  ($\times 10^{-3}$),
	]
\addplot[
smooth,
color=g-red,
mark=square*,
line width=1.5pt
]
coordinates {
(0,0.109)(1,0.195)(2,0.196)(3,0.168)(4,0.223)(5,0.215)(6,0.178)(7,0.225)(8,0.276)(9,0.285)(10,0.224)(11,0.157)(12,0.169)(13,0.216)(14,0.206)(15,0.205)(16,0.172)(17,0.192)(18,0.201)(19,0.215)(20,0.232)(21,0.186)(22,0.186)(23,0.215)(24,0.13)(25,0.181)(26,0.188)(27,0.2)(28,0.179)(29,0.18)(30,0.155)(31,0.172)(32,0.108)(33,0.152)(34,0.115)(35,0.093)(36,0.07)(37,0.074)(38,0.112)(39,0.117)(40,0.106)(41,0.122)(42,0.092)(43,0.127)(44,0.112)(45,0.137)(46,0.133)(47,0.146)(48,0.197)(49,0.206)(50,0.143)(51,0.25)(52,0.3)(53,0.311)(54,0.295)(55,0.324)(56,0.294)(57,0.336)(58,0.381)(59,0.343)(60,0.566)(61,0.555)(62,0.54)(63,0.559)(64,0.563)(65,0.638)(66,0.605)(67,0.652)(68,0.694)(69,0.622)(70,0.658)(71,0.622)(72,0.583)(73,0.598)(74,0.563)(75,0.637)(76,0.638)(77,0.585)(78,0.619)(79,0.651)(80,0.654)(81,0.616)(82,0.669)(83,0.641)(84,0.625)(85,0.594)(86,0.613)(87,0.64)(88,0.546)(89,0.567)(90,0.596)(91,0.621)(92,0.584)(93,0.609)(94,0.599)(95,0.518)(96,0.551)(97,0.538)(98,0.552)(99,0.499)(100,0.536)(101,0.564)(102,0.522)(103,0.53)(104,0.573)(105,0.568)(106,0.517)(107,0.584)(108,0.559)(109,0.498)(110,0.494)(111,0.579)(112,0.576)(113,0.578)
};\label{plot0}
\addplot[
smooth,
color=g-yellow,
mark=square*,
line width=1.5pt
]
coordinates {
(0,0.093)(1,0.08)(2,0.043)(3,0.06)(4,0.133)(5,0.105)(6,0.143)(7,0.126)(8,0.088)(9,0.029)(10,0.101)(11,0.039)(12,0.096)(13,0.059)(14,0.063)(15,0.032)(16,0.042)(17,0.057)(18,0.082)(19,0.038)(20,0.052)(21,0.06)(22,0.077)(23,0.083)(24,0.091)(25,0.092)(26,0.046)(27,0.086)(28,0.11)(29,0.089)(30,0.132)(31,0.093)(32,0.138)(33,0.077)(34,0.117)(35,0.104)(36,0.081)(37,0.094)(38,0.139)(39,0.133)(40,0.09)(41,0.091)(42,0.085)(43,0.115)(44,0.103)(45,0.153)(46,0.197)(47,0.158)(48,0.149)(49,0.156)(50,0.247)(51,0.291)(52,0.465)(53,0.547)(54,0.601)(55,0.635)(56,0.748)(57,0.927)(58,1.111)(59,1.154)(60,1.216)(61,1.319)(62,1.287)(63,1.363)(64,1.332)(65,1.294)(66,1.282)(67,1.307)(68,1.289)(69,1.24)(70,1.344)(71,1.358)(72,1.297)(73,1.302)(74,1.319)(75,1.253)(76,1.327)(77,1.344)(78,1.354)(79,1.368)(80,1.372)(81,1.425)(82,1.361)(83,1.347)(84,1.266)(85,1.241)(86,1.317)(87,1.258)(88,1.271)(89,1.309)(90,1.222)(91,1.244)(92,1.261)(93,1.172)(94,1.194)(95,1.222)(96,1.209)(97,1.194)(98,1.188)(99,1.218)(100,1.21)(101,1.175)(102,1.168)(103,1.171)(104,1.209)(105,1.171)(106,1.24)(107,1.234)(108,1.17)(109,1.215)(110,1.231)(111,1.15)(112,1.223)(113,1.187)
};\label{plot1}
\addplot[
smooth,
color=uw,
mark=square*,
line width=1.5pt
]
coordinates {
(0,0.085)(1,0.164)(2,0.077)(3,0.14)(4,0.091)(5,0.085)(6,0.12)(7,0.169)(8,0.148)(9,0.199)(10,0.119)(11,0.137)(12,0.176)(13,0.14)(14,0.182)(15,0.104)(16,0.13)(17,0.13)(18,0.131)(19,0.11)(20,0.133)(21,0.176)(22,0.135)(23,0.106)(24,0.089)(25,0.121)(26,0.087)(27,0.129)(28,0.183)(29,0.144)(30,0.104)(31,0.146)(32,0.144)(33,0.153)(34,0.16)(35,0.135)(36,0.168)(37,0.096)(38,0.144)(39,0.285)(40,0.349)(41,0.347)(42,0.543)(43,0.642)(44,0.722)(45,0.761)(46,0.868)(47,0.995)(48,0.993)(49,1.023)(50,0.94)(51,0.913)(52,0.956)(53,1.039)(54,0.989)(55,1.047)(56,1.106)(57,1.024)(58,1.022)(59,1.312)(60,1.362)(61,1.331)(62,1.236)(63,1.213)(64,1.252)(65,1.26)(66,1.227)(67,1.183)(68,1.166)(69,1.165)(70,1.204)(71,1.17)(72,1.169)(73,1.194)(74,1.142)(75,1.178)(76,1.186)(77,1.182)(78,1.196)(79,1.144)(80,1.216)(81,1.159)(82,1.164)(83,1.223)(84,1.204)(85,1.194)(86,1.254)(87,1.192)(88,1.155)(89,1.16)(90,1.208)(91,1.154)(92,1.177)(93,1.127)(94,1.184)(95,1.208)(96,1.153)(97,1.204)(98,1.128)(99,1.186)(100,1.209)(101,1.142)(102,1.089)(103,1.141)(104,1.188)(105,1.154)(106,1.097)(107,1.125)(108,1.109)(109,1.126)(110,1.177)(111,1.139)(112,1.146)(113,1.163)
};\label{plot2}
\addplot[
smooth,
color=g-grey,
mark=square*,
line width=1.5pt
]
coordinates {
(0,0.091)(1,0.126)(2,0.109)(3,0.132)(4,0.141)(5,0.203)(6,0.123)(7,0.202)(8,0.199)(9,0.163)(10,0.13)(11,0.185)(12,0.124)(13,0.152)(14,0.169)(15,0.119)(16,0.145)(17,0.122)(18,0.092)(19,0.101)(20,0.085)(21,0.068)(22,0.075)(23,0.104)(24,0.084)(25,0.169)(26,0.133)(27,0.141)(28,0.157)(29,0.157)(30,0.123)(31,0.195)(32,0.218)(33,0.303)(34,0.341)(35,0.526)(36,0.618)(37,0.639)(38,0.687)(39,0.855)(40,0.891)(41,1.017)(42,1.156)(43,1.218)(44,1.333)(45,1.301)(46,1.304)(47,1.375)(48,1.38)(49,1.328)(50,1.307)(51,1.298)(52,1.354)(53,1.424)(54,1.32)(55,1.421)(56,1.438)(57,1.529)(58,1.567)(59,1.643)(60,1.52)(61,1.573)(62,1.64)(63,1.778)(64,1.837)(65,1.734)(66,1.768)(67,1.825)(68,1.89)(69,1.848)(70,1.899)(71,1.841)(72,1.897)(73,1.898)(74,1.918)(75,1.887)(76,1.927)(77,1.929)(78,2.002)(79,1.94)(80,2.022)(81,1.984)(82,2.018)(83,2.054)(84,2.083)(85,2.069)(86,1.981)(87,2.064)(88,2.083)(89,2.096)(90,2.038)(91,2.036)(92,2.064)(93,2.073)(94,2.034)(95,2.036)(96,2.075)(97,2.031)(98,2.048)(99,2.111)(100,2.096)(101,2.048)(102,2.019)(103,2.037)(104,2.137)(105,2.087)(106,2.078)(107,2.08)(108,2.067)(109,2.078)(110,2.071)(111,2.149)(112,2.072)(113,2.045)
};\label{plot3}
\addplot[
smooth,
color=g-green,
mark=square*,
line width=1.5pt
]
coordinates {
(0,0.027)(1,0.086)(2,0.089)(3,0.114)(4,0.185)(5,0.15)(6,0.149)(7,0.166)(8,0.133)(9,0.145)(10,0.101)(11,0.159)(12,0.165)(13,0.168)(14,0.121)(15,0.117)(16,0.146)(17,0.092)(18,0.088)(19,0.142)(20,0.171)(21,0.115)(22,0.082)(23,0.143)(24,0.089)(25,0.145)(26,0.117)(27,0.167)(28,0.095)(29,0.131)(30,0.115)(31,0.136)(32,0.195)(33,0.128)(34,0.156)(35,0.176)(36,0.225)(37,0.192)(38,0.194)(39,0.223)(40,0.233)(41,0.236)(42,0.237)(43,0.273)(44,0.356)(45,0.395)(46,0.392)(47,0.419)(48,0.357)(49,0.379)(50,0.46)(51,0.378)(52,0.443)(53,0.379)(54,0.508)(55,0.483)(56,0.455)(57,0.471)(58,0.517)(59,0.432)(60,0.47)(61,0.455)(62,0.468)(63,0.527)(64,0.524)(65,0.533)(66,0.516)(67,0.525)(68,0.544)(69,0.629)(70,0.65)(71,0.621)(72,0.569)(73,0.634)(74,0.659)(75,0.695)(76,0.611)(77,0.626)(78,0.674)(79,0.686)(80,0.616)(81,0.638)(82,0.692)(83,0.675)(84,0.705)(85,0.732)(86,0.728)(87,0.728)(88,0.737)(89,0.883)(90,0.801)(91,0.785)(92,0.831)(93,0.799)(94,0.784)(95,0.848)(96,0.834)(97,0.866)(98,0.866)(99,0.843)(100,0.787)(101,0.86)(102,0.837)(103,0.857)(104,0.796)(105,0.834)(106,0.816)(107,0.833)(108,0.843)(109,0.824)(110,0.833)(111,0.857)(112,0.777)(113,0.794)
};\label{plot4}

	     \addlegendimage{/pgfplots/refstyle=plot0}        \addlegendentry{china}
    \addlegendimage{/pgfplots/refstyle=plot1}          \addlegendentry{lockdown}
    
     \addlegendimage{/pgfplots/refstyle=plot2}        \addlegendentry{trump}
     \addlegendimage{/pgfplots/refstyle=plot3}        \addlegendentry{hospitals}
     \addlegendimage{/pgfplots/refstyle=plot4}        \addlegendentry{increasing deaths and cases}

\end{axis}

\end{tikzpicture}
     \end{subfigure}

              \caption{Daily intensity scores for different subcategories for {\it anger}  on Twitter. }
              \label{anger}

\end{figure*}
\begin{figure*}[!ht]     
     \centering
     \begin{subfigure}{1\textwidth}
\begin{tikzpicture}
\begin{axis}[
    	   width=1\columnwidth,
	    height=0.4\columnwidth,
	    legend cell align=left,
	    legend style={at={(0.32, 0.55)},anchor=south east},
	    xtick={0,14,28,42,56,70,84,98,112},
	    xticklabels={Jan.20, Feb.3, Feb.17, Mar.2, Mar.16, Mar.30, Apr.13, Apr.27, May.11},
	    xticklabel style = {font=\small},
   		ymin=0, ymax=2.5,
   		xtick pos=left,
   		xtick align=outside,
	    xmin=1,xmax=120,
	    mark options={mark size=0.5},
		font=\small,
   	 	ymajorgrids=true,
    	xmajorgrids=true,
    	xlabel=Date,
        ylabel=Intensity Score  ($\times 10^{-3}$),
	]
\addplot[
smooth,
color=g-red,
mark=square*,
line width=1.5pt
]
coordinates {
(0,0.085)(1,0.108)(2,0.053)(3,0.105)(4,0.076)(5,0.111)(6,0.046)(7,0.052)(8,0.028)(9,0.043)(10,0.066)(11,0.12)(12,0.026)(13,0.037)(14,0.043)(15,0.074)(16,0.123)(17,0.106)(18,0.022)(19,0.101)(20,0.109)(21,0.105)(22,0.066)(23,0.05)(24,0.106)(25,0.114)(26,0.034)(27,0.015)(28,0.104)(29,0.044)(30,0.071)(31,0.099)(32,0.142)(33,0.113)(34,0.238)(35,0.176)(36,0.254)(37,0.29)(38,0.293)(39,0.339)(40,0.33)(41,0.295)(42,0.321)(43,0.342)(44,0.337)(45,0.475)(46,0.48)(47,0.545)(48,0.61)(49,0.635)(50,0.69)(51,0.705)(52,0.783)(53,0.955)(54,0.988)(55,1.115)(56,1.082)(57,1.131)(58,1.19)(59,1.29)(60,1.348)(61,1.36)(62,1.499)(63,1.599)(64,1.545)(65,1.635)(66,1.637)(67,1.658)(68,1.808)(69,1.705)(70,1.715)(71,1.741)(72,1.815)(73,1.813)(74,1.875)(75,1.913)(76,1.852)(77,1.855)(78,1.875)(79,1.852)(80,1.937)(81,1.998)(82,1.899)(83,1.895)(84,1.873)(85,1.928)(86,1.954)(87,1.894)(88,1.872)(89,1.973)(90,1.847)(91,1.949)(92,1.928)(93,1.943)(94,1.881)(95,1.956)(96,2.035)(97,1.911)(98,2.039)(99,1.962)(100,1.92)(101,1.877)(102,1.879)(103,1.969)(104,2.008)(105,1.91)(106,1.836)(107,1.93)(108,1.893)(109,1.912)(110,1.883)(111,1.881)(112,1.885)(113,2.01)
};\label{plot0}
\addplot[
smooth,
color=g-yellow,
mark=square*,
line width=1.5pt
]
coordinates {
(0,0.108)(1,0.097)(2,0.071)(3,0.119)(4,0.043)(5,0.039)(6,0.114)(7,0.111)(8,0.109)(9,0.037)(10,0.108)(11,0.094)(12,0.081)(13,0.118)(14,0.076)(15,0.127)(16,0.051)(17,0.051)(18,0.065)(19,0.147)(20,0.144)(21,0.121)(22,0.099)(23,0.074)(24,0.154)(25,0.134)(26,0.081)(27,0.071)(28,0.086)(29,0.135)(30,0.081)(31,0.114)(32,0.105)(33,0.187)(34,0.187)(35,0.22)(36,0.23)(37,0.19)(38,0.255)(39,0.246)(40,0.288)(41,0.241)(42,0.282)(43,0.311)(44,0.361)(45,0.307)(46,0.392)(47,0.47)(48,0.505)(49,0.447)(50,0.587)(51,0.663)(52,0.686)(53,0.853)(54,1.086)(55,1.231)(56,1.282)(57,1.297)(58,1.381)(59,1.47)(60,1.472)(61,1.426)(62,1.542)(63,1.584)(64,1.607)(65,1.704)(66,1.747)(67,1.796)(68,1.777)(69,1.971)(70,1.977)(71,2.047)(72,1.932)(73,1.951)(74,1.971)(75,1.994)(76,2.1)(77,2.014)(78,1.991)(79,2.052)(80,1.919)(81,2.073)(82,2.037)(83,1.994)(84,1.985)(85,1.961)(86,1.979)(87,2.031)(88,2.06)(89,2.03)(90,2.019)(91,2.048)(92,1.998)(93,1.925)(94,1.965)(95,1.932)(96,1.897)(97,1.954)(98,2.03)(99,2.068)(100,2.131)(101,2.157)(102,2.169)(103,2.115)(104,2.093)(105,2.092)(106,2.144)(107,2.019)(108,2.113)(109,2.113)(110,2.049)(111,2.095)(112,2.113)(113,2.072)
};\label{plot1}
\addplot[
smooth,
color=uw,
mark=square*,
line width=1.5pt
]
coordinates {
(0,0.168)(1,0.115)(2,0.126)(3,0.084)(4,0.129)(5,0.11)(6,0.125)(7,0.127)(8,0.126)(9,0.122)(10,0.161)(11,0.194)(12,0.217)(13,0.229)(14,0.166)(15,0.22)(16,0.212)(17,0.132)(18,0.146)(19,0.245)(20,0.187)(21,0.173)(22,0.156)(23,0.115)(24,0.179)(25,0.126)(26,0.122)(27,0.158)(28,0.129)(29,0.089)(30,0.143)(31,0.186)(32,0.19)(33,0.2)(34,0.261)(35,0.413)(36,0.563)(37,0.682)(38,0.7)(39,0.784)(40,0.737)(41,0.83)(42,0.8)(43,0.865)(44,0.918)(45,1.004)(46,1.157)(47,1.175)(48,1.298)(49,1.492)(50,1.471)(51,1.419)(52,1.547)(53,1.504)(54,1.333)(55,1.445)(56,1.387)(57,1.511)(58,1.369)(59,1.433)(60,1.455)(61,1.475)(62,1.566)(63,1.489)(64,1.408)(65,1.43)(66,1.379)(67,1.436)(68,1.448)(69,1.327)(70,1.439)(71,1.376)(72,1.41)(73,1.323)(74,1.472)(75,1.312)(76,1.505)(77,1.42)(78,1.468)(79,1.458)(80,1.419)(81,1.413)(82,1.457)(83,1.432)(84,1.506)(85,1.472)(86,1.514)(87,1.449)(88,1.476)(89,1.471)(90,1.4)(91,1.47)(92,1.396)(93,1.414)(94,1.454)(95,1.366)(96,1.414)(97,1.409)(98,1.471)(99,1.434)(100,1.426)(101,1.446)(102,1.375)(103,1.461)(104,1.369)(105,1.414)(106,1.386)(107,1.428)(108,1.418)(109,1.433)(110,1.509)(111,1.38)(112,1.426)(113,1.329)
};\label{plot2}
\addplot[
smooth,
color=g-grey,
mark=square*,
line width=1.5pt
]
coordinates {
(0,0.066)(1,0.091)(2,0.118)(3,0.106)(4,0.102)(5,0.099)(6,0.098)(7,0.117)(8,0.102)(9,0.106)(10,0.099)(11,0.057)(12,0.124)(13,0.101)(14,0.134)(15,0.127)(16,0.114)(17,0.048)(18,0.118)(19,0.144)(20,0.088)(21,0.06)(22,0.1)(23,0.116)(24,0.146)(25,0.159)(26,0.097)(27,0.166)(28,0.164)(29,0.095)(30,0.145)(31,0.083)(32,0.084)(33,0.088)(34,0.072)(35,0.098)(36,0.145)(37,0.101)(38,0.122)(39,0.096)(40,0.127)(41,0.103)(42,0.141)(43,0.131)(44,0.208)(45,0.211)(46,0.18)(47,0.178)(48,0.208)(49,0.288)(50,0.242)(51,0.298)(52,0.326)(53,0.4)(54,0.525)(55,0.536)(56,0.586)(57,0.729)(58,0.833)(59,0.812)(60,0.958)(61,1.007)(62,1.077)(63,1.114)(64,1.218)(65,1.142)(66,1.183)(67,1.329)(68,1.36)(69,1.353)(70,1.329)(71,1.42)(72,1.363)(73,1.363)(74,1.431)(75,1.352)(76,1.326)(77,1.446)(78,1.479)(79,1.438)(80,1.485)(81,1.614)(82,1.452)(83,1.539)(84,1.51)(85,1.507)(86,1.477)(87,1.586)(88,1.558)(89,1.618)(90,1.672)(91,1.633)(92,1.625)(93,1.651)(94,1.595)(95,1.61)(96,1.686)(97,1.739)(98,1.74)(99,1.652)(100,1.712)(101,1.577)(102,1.643)(103,1.72)(104,1.73)(105,1.714)(106,1.637)(107,1.706)(108,1.757)(109,1.768)(110,1.777)(111,1.79)(112,1.764)(113,1.786)
};\label{plot3}
\addplot[
smooth,
color=g-green,
mark=square*,
line width=1.5pt
]
coordinates {
(0,0.078)(1,0.136)(2,0.147)(3,0.116)(4,0.084)(5,0.077)(6,0.146)(7,0.136)(8,0.135)(9,0.068)(10,0.067)(11,0.096)(12,0.106)(13,0.062)(14,0.101)(15,0.067)(16,0.139)(17,0.124)(18,0.12)(19,0.077)(20,0.079)(21,0.151)(22,0.139)(23,0.089)(24,0.123)(25,0.113)(26,0.092)(27,0.119)(28,0.154)(29,0.095)(30,0.1)(31,0.185)(32,0.096)(33,0.154)(34,0.114)(35,0.172)(36,0.123)(37,0.161)(38,0.151)(39,0.156)(40,0.265)(41,0.3)(42,0.312)(43,0.392)(44,0.339)(45,0.347)(46,0.344)(47,0.347)(48,0.362)(49,0.429)(50,0.367)(51,0.386)(52,0.418)(53,0.376)(54,0.428)(55,0.436)(56,0.44)(57,0.436)(58,0.416)(59,0.404)(60,0.439)(61,0.42)(62,0.456)(63,0.47)(64,0.45)(65,0.434)(66,0.469)(67,0.465)(68,0.437)(69,0.499)(70,0.509)(71,0.495)(72,0.464)(73,0.45)(74,0.535)(75,0.52)(76,0.508)(77,0.471)(78,0.464)(79,0.459)(80,0.508)(81,0.466)(82,0.527)(83,0.458)(84,0.475)(85,0.523)(86,0.459)(87,0.459)(88,0.449)(89,0.535)(90,0.508)(91,0.443)(92,0.465)(93,0.46)(94,0.448)(95,0.491)(96,0.528)(97,0.439)(98,0.454)(99,0.446)(100,0.512)(101,0.461)(102,0.502)(103,0.504)(104,0.479)(105,0.438)(106,0.506)(107,0.467)(108,0.464)(109,0.453)(110,0.485)(111,0.454)(112,0.484)(113,0.432)
};\label{plot4}

	     \addlegendimage{/pgfplots/refstyle=plot0}        \addlegendentry{syndrome and being infected}
    \addlegendimage{/pgfplots/refstyle=plot1}          \addlegendentry{families}
    
     \addlegendimage{/pgfplots/refstyle=plot2}        \addlegendentry{finance and economy}
     \addlegendimage{/pgfplots/refstyle=plot3}        \addlegendentry{jobs and food}
     \addlegendimage{/pgfplots/refstyle=plot4}        \addlegendentry{increasing deaths and cases}

\end{axis}

\end{tikzpicture}
     \end{subfigure}

              \caption{Daily intensity scores for different subcategories for {\it worry}  on Twitter. }
              \label{worry}

\end{figure*}
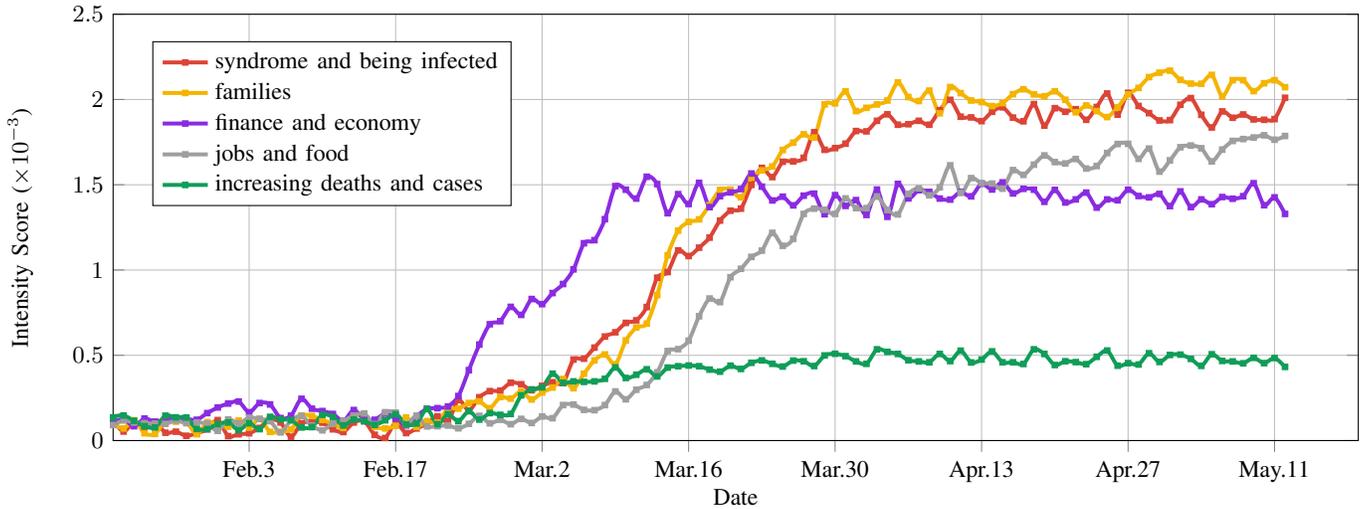

\subsection{Clustering Trigger Mentions}
Since different extracted tokens may refer to the same concept or topic, we would like to cluster the extracted trigger mentions. 
The method of supervised classification is unsuitable for this purpose since (1) it is hard to predefine a list of potential triggers to people's anger or worry; (2) it is extremely labor-intensive to annotate tweets with worry types or anger types and (3) these types may change over time. 
For these reasons, we decided to use semi-supervised approaches that will automatically induce worry/anger types that match the data. 
We adopt an approach based on LDA \cite{blei2003latent}. It was inspired by work on unsupervised information extraction \cite{chambers2011template,ritter2012open,li2014major}. 

We use the emotion {\it anger} to illustrate how trigger mentions are clustered. 
Each extracted trigger mentions for {\it anger} is modeled as a mixture of {\it anger} types.
Here we use subcategory, type, and topic interchangeablely, all referring to the cluster of similar mentions. 
 Each topic is characterized by a distribution over triggers, in addition to a distribution over dates on which a user talks about the topic. 
Taking dates into account encourages triggers that are mentioned on the same date to be assigned to the same topic.  
We used collapsed Gibbs Sampling \cite{griffiths2004finding} for inference. 
For each emotion, we ran Gibbs Sampling with 20 topics for 1,000 iterations, obtaining the hidden variable assignments in the last iteration. 
Then we manually inspected the top mentions for different topics and abandoned the incoherent ones. 
 
The daily  intensity score for a given subcategory $k$ belonging to emotion $y$ is given as follows: 
 \begin{equation}
S(t,y, k) = \frac{1}{|X_t|} \sum_{x\in X_t} p (k|x)  {\bf I}(y_x = y)
\end{equation}
where $p(k|x)$ is computed based on the parameters of the latent variable model.

 \subsection{Analyses}
We report the top triggers for different emotions in Table \ref{top-entity}.
 We adopt a simple strategy of reporting the most frequent triggers for different emotions. 
For {\it sadness}, the most frequent triggering events and topics are being test positive, and the death of families and friends. 
For {\it anger}, the top triggers are shutdown, quarantine and other mandatory rules. 
People also express their anger towards
public figures such as
 President Donald Trump,
Mike Pence, 
 along with China and Chinese. 
For {\it worry}, the top triggers include jobs, getting the virus, payments and families.
For {\it happiness}, the top triggers are recovering from the disease, city reopening and returning to work. 
For {\it surprise}, the public are mostly surprised by the virus itself, its spread and the mass deaths it caused. 

Next we report the results of the mention clustering for {\it anger} and {\it worry} in Tables \ref{anger} and \ref{worry},  respectively.
The unsupervised clustering reveals clearer patterns in the triggering events:  top subcategories for {\it anger} include  China with  racist words such as {\it chink} and {\it chingchong};  lockdown and social distancing; 
public figures like President Donal Trump and Mike Pence; 
treatments in hospitals, and the increasing cases and deaths;
Table \ref{anger} displays the change of intensity scores for the subcategories of {\it anger}.
We observe a sharp increase in public anger toward China and Chinese around March 20th, 
in coincidence with President Donald Trump calling coronavirus 'Chinese virus' in his tweets. 
Public anger towards the lockdown sharply escalated in mid-March, but decreased a bit after late April when some places started to reopen.

Top subcategories for {\it worry} include syndromes for COVID-19, finance and economy, families, jobs and food, and increasing cases and deaths. 
  Table \ref{worry} displays the change of intensity scores for subcategories of {\it worry}.
People increasingly worried about families over time. 
It is interesting to see that the worry about finance and economy started going up in mid-February, earlier than other subcategories.   

\section{Conclusion}
In this paper, we 
perform analyses on 
 topic trends, sentiments and emotions of the public in the time of COVID-19 on social media.
By tracking the change of public emotions over time, 
our work reveals how the general public reacts to different events and government policy. 
Our study provides a computational approach to understanding public affect towards the pandemic in real-time, and could help create better solutions and interventions
for fighting this global crisis.  

\bibliography{covid}
\bibliographystyle{acl_natbib}

%

\end{document}